\documentclass[12pt]{JHEP3}

\usepackage{amsmath}

\newcommand{\be}{\begin{eqnarray}}
\newcommand{\ee}{\end{eqnarray}}

\newcommand{\bn}{\begin{enumerate}}
\newcommand{\en}{\end{enumerate}}

\def\Tr{{\rm Tr}}

\title{\bf Duality between ${\cal N}=5$ and ${\cal N}=6$ Chern-Simons matter theory}

\author{ Sangmo Cheon $^{1}$, Dongmin Gang $^{2}$, Chiung Hwang $^{3}$, Satoshi Nagaoka $^{4}$,  Jaemo Park $^{3,5}$

\\

$^{1}$ Department of Physics and Astronomy \& Center for Theoretical
Physics,
 Seoul National University, Seoul 151-747, Korea
\\
$^{2}$ Korea Institute for Advanced Study, Seoul,130-012,  Korea
\\
$^{3}$Department of Physics, POSTECH, Pohang 790-784, Korea
\\
$^{4}$Asia Pacific Center for Theoretical Physics, Pohang, 790-784,
Korea
\\
$^{5}$Postech Center for Theoretical Physics (PCTP), Postech, Pohang
  790-784, Korea

\\
\\
E-mail: \email{sangmocheon@gmail.com,
 arima275@kias.re.kr, c\_hwang@postech.ac.kr,
nagaoka@apctp.org, jaemo@postech.ac.kr} } %%
\abstract{We provide evidences for the duality between  ${\cal N}=6$
$U(M)_{4} \times U(N)_{-4}$ Chern-Simons matter theory and ${\cal
N}=5$ $O(\hat{M})_{2} \times USp(2\hat{N})_{-1}$ theory for a
suitable $\hat{M},\hat{N}$ by working out the superconformal index,
which shows perfect matching. For ${\cal N}=5$ theories, we  show
that supersymmetry is enhanced to ${\cal N}=6$ by explicitly
constructing monopole operators filling in $SO(6)_R$ $R$-currents.
Finally we work out the large $N$ index of $O(2N)_{2k} \times
USp(2N)_{-k}$ and show that it exactly matches with the gravity
index on $AdS_4 \times S^7/D_k$, which further provides additional
evidence for the duality between the ${\cal N}=5$ and ${\cal N}=6$
theory for $k=1$. }

\preprint{APCTP Pre2012-014\\KIAS-P12050}

\begin{document}

\section{Introduction}

There have been much progress in understanding 3-d superconformal
field theories recently. Many of 3-d superconformal field theories
are realized as supersymmetric Chern-Simons matter (SCSM) theories.
Quite exhaustive classes of ${\cal N}\geq 4$ SCSM theories are constructed
in \cite{Gustavsson:2007vu,Bagger:2007jr,Gaiotto:2008sd,Hosomichi:2008jd,Aharony:2008ug,
Hosomichi:2008jb,Imamura:2008dt}. Among them, the most famous example is ABJM theory, ${\cal N}=6$
$U(N)_k \times U(N)_{-k}$ Chern-Simons matter theory describing M2
branes on $C^4/Z_k$. The curious fact of ABJM theory is that
supersymmetry is enhanced to ${\cal N}=8$ for $k=1,2$. This was discussed at
\cite{Aharony:2008ug,Benna:2009xd,Gustavsson:2009pm,Kwon:2009ar} and further clarified
where the monopole operators filling ${\cal N}=8$
$R$-currents are explicitly constructed \cite{Bashkirov:2010kz}.

Another interesting feature of SCSM theories is that 3-d analogue of
Seiberg-duality holds for some classes of the theories. For example
there are various evidences that ${\cal N}=2$ $U(N)_k$ SCSM theory
with $N_f$ fundamental flavors is dual to ${\cal N}=2$
$U(N_f-N+|k|)_{-k}$ with $N_f$ fundamental flavors with additional
meson fields \cite{Giveon:2008zn,Willett:2011gp,Hwang:2011qt}. While
in 4-d Seiberg dualities hold for ${\cal N}=1$ supersymmetric
theories, there are several evidences that such duality holds for
SCSM theories with ${\cal N}\geq 2$ theories \cite{seok2012, kapustin2010}.

In these respects, we had better explore further models, which would
enrich our understanding of 3-d SCSM theories. In this note we are
interested in another type of dual pairs of SCSM theory in 3-d. The
theories we are interested in are ${\cal N}=6$ $U(M)_{4} \times
U(N)_{-4}$ and ${\cal N}=5$ $O(\hat{M})_{2} \times
USp(2\hat{N})_{-1}$. There is a conjecture that this pair is dual to
each other for suitable $\hat{M}, \hat{N}$ for given $M$, $N$ as
will be explicitly stated in the main text. This duality is
interesting in several ways. Firstly, ${\cal N}=6$ $U(N+l)_4 \times
U(N)_{-4}$ Chern-Simons theory itself exhibits Seiberg-like
dualities, i.e., it is dual to $U(N)_4\times U(N+4-l)_{-4}$ with
$l\leq 4$. And the similar holds for ${\cal N}=5$ theory as well. In
addition, there is a separate duality connecting ${\cal N}=6$ theory
and ${\cal N}=5$ theory, whose physical origin is not clear at this
point.
% We provide evidences for Seiberg-like dualities among
%suitable ${\cal N}=6$ theories and ${\cal N}=5$ theories by working
%out superconformal index at section 2.
%Thus we provide evidence for
%Seiberg-like dualities for product gauge groups with higher
%supersymmetry for a special case.
Index computation provides evidence that ${\cal N}=6$ $U(M)_{4}
\times U(N)_{-4}$ theory is dual to ${\cal N}=5$ $O(\hat{M})_{2}
\times USp(2\hat{N})_{-1}$ theory. As a byproduct of the
computation, we can provide evidences for Seiberg-like dualities
among  ${\cal N}=6$ theories and among ${\cal N}=5$ theories for
simple cases. Secondly, since ${\cal N}=5$ theory is dual to ${\cal
N}=6$ theory for a particular choice of Chern-Simons level, the
supersymmetry of ${\cal N}=5$ theory should be extended to ${\cal
N}=6$. Adopting the method of \cite{Bashkirov:2010kz}, we explicitly
construct the monopole operators filling in $Spin(6)$ $R$-currents,
thereby showing the supersymmetry is indeed enhanced at section 3.
Finally we work out the large $N$ limit of the superconformal index of
${\cal N}=5$ $O(2N)_{2k} \times USp(2N)_{-k}$ theory and matches it
to the gravity index of $AdS_4 \times S^7/D_k$ with $D_k$ dihedral
group of $4k$ elements. This not only provides evidence for the
conjecture that ${\cal N}=5$ theory has the gravitational dual of $AdS_4
\times S^7/D_k$ but also provides additional evidence for the
equivalence between the ${\cal N}=5$ $O(2N)_{2} \times USp(2N)_{-1}$
theory and ${\cal N}=6$ $U(N)_4 \times U(N)_{-4}$ theory, which is
in turn dual to the gravity theory on $AdS_4 \times S^7/Z_4$ since
$D_1=Z_4$. In appendix we provide the details of the index formulae
used in the main text.

\section{Superconformal index}
Let us first discuss the structure of ${\cal N}=6$ $U(M)_k\times
U(N)_{-k}$ theory, and  the ${\cal N}=5$ $O(M)_{2k}\times
USp(2N)_{-k}$ theory, which we abbreviate $U(M|N)$, $OSp(M|2N)$
respectively.\footnote{In our convention, $SO(2)_1$ Chern-Simons
theory is equivalent to $U(1)_1$ theory.} The ${\cal N}=6$ $U(M|N)$
theory contains four superfields $C_I$ in the fundamental
representation $\bf 4$ of the $R$-symmetry $SU(4)_R\simeq Spin(6)_R$
and in the bifundamental representation of the gauge group. The
superfields $C_I$ can be written in the ${\cal N}=2$ formalism as
$C_I=(A_1,A_2,\bar B^{\dot 1}, \bar B^{\dot 2})$ where $A_a$ and
$B_{\dot b}$ are four ${\cal N}=2$ chiral multiplets in the
representation $({\bf 2},{\bf 1})$ and $({\bf 1},{\bf 2})$ of
$SU(2)_A\times SU(2)_B\subset SU(4)_R$ respectively. The theory has
the superpotential
\begin{equation}
W=-\frac{2\pi}{k}\epsilon^{ab}\epsilon^{\dot a\dot b}{\rm
tr}\left(A_aB_{\dot a}A_bB_{\dot b}\right)=\frac{4\pi}{k}{\rm
tr}\left(A_1B_2A_2B_1 - A_1B_2A_1B_2\right).
\end{equation}
The ${\cal N}=5$ $OSp(M|2N)$ theory also contains the superfields
$C_I=(A_1,A_2,\bar B^{\dot 1},\bar B^{\dot 2})$ with identifications
$A_1=B_{\dot 1}^TJ$ and $A_2=B_{\dot 2}^TJ$ where $J$ is the
invariant antisymmetric matrix of the symplectic group. These
identifications preserve $USp(4)_R\subset SU(4)_R$ such that the
$R$-symmetry of the theory is $USp(4)_R\simeq Spin(5)_R$. The theory
has the superpotential
\begin{equation}
W=\frac{4\pi}{k}{\rm tr}\left(A_1JA^T_2 A_2JA^T_1 - A_1JA^T_1 A_2JA^T_2\right).
\end{equation}

Curiously, it seems that there exist a duality between these two
theories when the Chern-Simons levels are $(k,-k)=(4,-4)$ for the
first theory and $(2k,-k)=(2,-1)$ for the second theory. More
specifically, the following dual relations were conjectured
\cite{Aharony:2008gk}:
\begin{eqnarray*}
 \begin{array}{r}
U(N)_4\times U(N)_{-4}
\end{array}&\leftrightarrow&\begin{array}{l}
O(2N)_2\times USp(2N)_{-1}
\end{array}\\
\begin{array}{r}
U(N+2)_4\times U(N)_{-4}
\end{array}&\leftrightarrow&\begin{array}{l}
O(2N+2)_2\times USp(2N)_{-1}
\end{array}\\
\left.\begin{array}{r}
U(N+1)_4\times U(N)_{-4}\\
U(N+3)_4\times U(N)_{-4}
\end{array}\right\}&\leftrightarrow&\left\{\begin{array}{l}
O(2N+1)_2\times USp(2N)_{-1}\\
USp(2N)_1\times O(2N+1)_{-2}
\end{array}\right.
\end{eqnarray*}
In fact, at \cite{Aharony:2008gk} the duality for the  first two
theories are not precisely stated and we improve it in our
paper.\footnote{At \cite{Aharony:2008gk}, it's claimed that the
first two theories on the left hand side are mapped to the first two
theories on the right hand side although which to which is not
clear.} The last pair on the left hand side are mapped to the last
pair on the right hand side. Furthermore, these two theories on each
side also are related by Seiberg-like duality. This is clear from
the brane construction of \cite{Aharony:2008gk}, where Seiberg-like
duality can be inferred from the NS-5 brane movement past the
infinite coupling.\footnote{More precisely, for example,
$U(N+1)_4\times U(N)_{-4}$ is equivalent to $U(N+3)_4 \times
U(N)_{-4}$ if we combine Seiberg-like duality and parity
transformation. } In the ${\cal N}=2$ setting, the Seiberg-like
dualities for product gauge groups are discussed in
\cite{Agarwal:2012wd}. See \cite{Benini} also related discussions on
3-d Seiberg-like dualities.

As shown at \cite{Aharony:2008gk}, all these theories have moduli space as the
symmetric product of $C^4/Z_4$. This is the first evidence for the
conjectured duality. In order to provide further evidences for the
claimed dualities, we resort to the superconformal index
computation. Let us discuss the general structures of the index. We
consider the superconformal index for 3-d $\mathcal{N}=2$
superconformal field theory (SCFT). The superconformal index for a higher
supersymmetric theory can be defined using their $\mathcal{N}=2$
subalgebra. The bosonic subgroup of the 3-d $\mathcal{N}=2$
superconformal algebra is $SO(2,3) \times SO(2) $. There are three
Cartan elements denoted by $\epsilon$, $j$ and $R$ which come from
three factors $SO(2)_\epsilon \times SO(3)_{j}\times SO(2)_R $ in
the bosonoic subalgebra.  One can define the superconformal index
for 3-d $\mathcal{N}=2$ SCFT as follows \cite{Bhattacharya:2008bja},
\begin{equation}
I=\Tr (-1)^F \exp (-\beta'\{Q, S\}) x^{\epsilon+j}y_j^{F_j}
\label{def:index}
\end{equation}
where $Q$ is a   supercharge with quantum numbers $\epsilon =
\frac{1}2$, $j = -\frac{1}{2}$ and $R=1$ and $S= Q^\dagger$. They
satisfy following anti-commutation relation:
\begin{equation}
 \{Q, S\}=\epsilon-R-j : = \Delta.
\end{equation}
In the index formula, the trace is taken over gauge-invariant local
operators in the SCFT defined on $\mathbb{R}^{1,2}$ or over states
in the SCFT on $\mathbb{R}\times S^2$. As is usual for Witten index
, only BPS states satisfying the bound $\Delta =0 $ contributes to
the index and the index is independent of $\beta'$. If we have
additional conserved charges commuting with chosen supercharges
($Q,S$), we can turn on the associated chemical potentials and the
index counts the number of BPS states with the specified quantum
number of the conserved charges denoted by $F_j$ in eq.
(\ref{def:index}). We simply set $y_j=1$ in the subsequent
computation.

 The index can be exactly calculated using the localization
technique, deforming the action by a $Q$-exact term and making the
Gaussian approximation exact \cite{Kim:2009wb,Imamura:2011su}. The
deformation we adopt  breaks the $R$-symmetry from $Spin({\cal
N})_R$ to $Spin({\cal N}-2)\times SO(2)_R$. For the details, readers
may refer to the appendix and \cite{Hwang:2011qt,Kim:2009wb}. In the
subsequent computation for confirming dualities, it is crucial that
we consider $O(N)$ theory instead of $SO(N)$ for the $OSp(N|2M)$
theory.

  Let us first consider the simplest case $U(1|1)$ theory. The index for the
$U(1|1)$ theory is given by
\begin{eqnarray}
I_{U(1|1)}(x)&=&1+4 x+11 x^2+12 x^3+25 x^4+12 x^5+44 x^6+24 x^7\nonumber\\
&&+32 x^8+{\cal O}\left(x^9\right).
\end{eqnarray}
On the other hand, the indices for the $OSp(2|2)$ theory and the $OSp(4|2)$ theory are given by
\begin{eqnarray}
I(x)_{OSp(2|2)}&=&1+4 x+11 x^2+12 x^3+25 x^4+12 x^5+44 x^6+24 x^7\\
&&+32 x^8+{\cal O}\left(x^9\right),\nonumber\\
I(x)_{OSp(4|2)}&=&1+4 x+12 x^2+8 x^3+27 x^4+36 x^5-36 x^6+{\cal O}\left(x^7\right).
\end{eqnarray}
The $U(1|1)$ theory and the $OSp(2|2)$ theory have the same index
while the $OSp(4|2)$ theory doesn't. Therefore, we provide the
evidence that the $U(1|1)$ theory and the $OSp(2|2)$ theory are  a
dual pair while one can expect that the $U(3|1)$ theory and the
$OSp(4|2)$ theory are another dual pair. The index for the $U(3|1)$
theory is given by
\begin{equation}
I_{U(3|1)}(x)=1+4 x+12 x^2+8 x^3+27 x^4+{\cal O}\left(x^5\right),
\end{equation}
which is the same as that for the $OSp(4|2)$ theory as expected, up
to the order of $x^4$. In a similar manner, we tested the duality
for some low rank cases. The result is given in Table~\ref{SCI}.
\begin{table}
\begin{tabular}{c|l}
\hline
$\begin{array}{c}
U_4\times U_{-4}\\
O_2\times USp_{-1}
\end{array}$&The superconformal index\\
\hline
\hline
$U(1|1)$&$1+4 x+11 x^2+12 x^3+25 x^4+12 x^5+44 x^6+24 x^7+32 x^8+{\cal O}\left(x^9\right)$\\
$OSp(2|2)$&$1+4 x+11 x^2+12 x^3+25 x^4+12 x^5+44 x^6+24 x^7+32 x^8+{\cal O}\left(x^9\right)$\\
\hline
$U(3|1)$&$1+4 x+12 x^2+8 x^3+27 x^4+{\cal O}\left(x^5\right)$\\
$OSp(4|2)$&$1+4 x+12 x^2+8 x^3+27 x^4+36 x^5-36 x^6+{\cal O}\left(x^7\right)$\\
\hline
$U(2|2)$&$1+4 x+22 x^2+56 x^3+131 x^4+252 x^5+516 x^6+{\cal O}\left(x^7\right)$\\
$OSp(4|4)$&$1+4 x+22 x^2+56 x^3+131 x^4+{\cal O}\left(x^5\right)$\\
\hline
$U(3|2)$&$1+4 x+22 x^2+60 x^3+134 x^4+200 x^5+556 x^6+{\cal O}\left(x^7\right)$\\
$OSp(5|4)$&$1+4 x+22 x^2+60 x^3+134 x^4+200 x^5+556 x^6+{\cal O}\left(x^7\right)$\\
\hline
$U(2|1)$&$1+4 x+12 x^2+8 x^3+27 x^4+32 x^5-20 x^6+128 x^7-65 x^8+{\cal O}\left(x^9\right)$\\
$U(4|1)$&$1 + 4 x + 12 x^2+{\cal O}\left(x^3\right)$\\
$\begin{array}{c}
OSp(3|2)/\\
USp(2)_1\times O(3)_{-2}
\end{array}$&$1+4 x+12 x^2+8 x^3+27 x^4+32 x^5-20 x^6+128 x^7-65 x^8+{\cal O}\left(x^9\right)$\\
\hline
\end{tabular}
\caption{The superconformal indices for some low rank dual pairs.}\label{SCI}
\end{table}

One can explicitly examine the computation result of the index in
detail. For example, the result indicates that every dual pair has
four gauge-invariant operators with $\epsilon+j=1$. By
state-operator correspondence of the conformal field theory, gauge
invariant operators on $\mathbb R^3$ have corresponding states in $\mathbb R \times
S^2$. These states can be easily found and are given as follows:
\begin{equation}
\bar A^{a}\bar B^{\dot b}\left|0,\cdots;0,\cdots\right>
\label{states1}
\end{equation}
for the $U(M|N)$ theory and
\begin{eqnarray}
&&\bar A^{[a}\bar A^{b]}\left|0,\cdots;0,\cdots\right>,\label{states2}\\
&&\bar A^{(a}_{1\hat 1} \bar A^{b)}_{1\hat 1}\left|1,0,\cdots;1,0,\cdots\right>\label{states3}
\end{eqnarray}
for the $OSp(M|2N)$ theory where $\left|m,n,\cdots;\tilde m,\tilde
n,\cdots\right>$ is a bare monopole state. The flavor indices $a,b$
and $\dot b$ run over $1, 2$. The omitted gauge indices for
\eqref{states1} and \eqref{states2} are properly contracted to form
gauge invariant states. The expression for the gauge indices of
\eqref{states3} is schematic. The expression means that the matter
fields are excited satisfying the Gauss law constraint such that the
states having nonvanishing GNO charges are gauge invariant. These
states exist regardless of the rank of the gauge group. The
existence of such states is crucial to examine the supersymmetry
enhancement in the next section.

\section{Supersymmetry enhancement}
The relevant facts about the ${\cal N}\geq 4$ superconformal algebra
are explained at \cite{Bashkirov:2011fr}.\footnote{See also
\cite{Gustav} for related but incomplete discussion of the enhanced
supersymmetry for ${\cal N}=5$ theory. } We are interested in the
stress tensor multiplet. The lowest component of the stress tensor
multiplet is an $SO(3)$ scalar and an antisymmetric rank 4 tensor of
$Spin({\cal N})_R$ with the conformal dimension $\epsilon=1$, where
$SO(3)$ denotes the rotation group of $\mathbb R^3$. Another component of
our interest is $R$-current, which is antisymmetric rank 2 $Spin({\cal
N})_R$ tensor with the conformal dimension 2. Starting from one
component, one obtains another component of the different conformal
dimension by acting supercharges and its conjugate. For example, $R$-currents can be obtained
from the scalar components by acting supercharges twice so that the  $R$-current
components have spin 1 with respect to $SO(3)$.

If there is a conserved global current, it belongs to a
supermultiplet. The lowest component of this supermultiplet is an
$SO(3)$ scalar which is an antisymmetric 2nd rank tensor of $Spin({\cal N})_R$
with $\epsilon=1$. Note that for ${\cal N}=6$, the lowest component
of the stress tensor multiplet and that of the global current multiplet are on the
same representation of $Spin(6)_R$ with the same conformal dimension.
It is indeed argued that this global part is a part of the stress
tensor multiplet so that every ${\cal N}=6$ has the global $U(1)$
symmetry.

Now we would like to argue that $R$-symmetry $Spin(5)_R$ is enhanced
to $Spin(6)_R$ for the special case of ${\cal N}=5$ theory we
discussed before so that it has ${\cal N}=6$. The strategy is to
look for the lowest scalar component of the stress tensor multiplet
then obtain the needed $R$-currents by acting superconformal
generators on the scalar component.

Let us consider the lowest scalar component. For ${\cal N}=6$ this is the
rank 4 antisymmetric tensor representation of $Spin(6)_R$ , $\bf
15$. It decomposes under $Spin(5)_R\subset Spin(6)_R$ as ${\bf
15}={\bf 5}\oplus{\bf 10}$ where $\bf 5$ and $\bf 10$ are
respectively rank 4 and rank 3 antisymmetric tensor
representations of $Spin(5)_R$. $\bf 5$ is the representation of the
lowest scalar component of the $Spin(5)_R$ stress tensor multiplet.
As explained at the previous section, we adopted the deformation
 breaking the $Spin(5)_R$ $R$-symmetry down to $Spin(3)\times SO(2)_R\simeq
SU(2)\times U(1)_R$, under which the 3rd rank tensor
representation $\bf 10$ of $Spin(5)_R$ decomposes as ${\bf 10}={\bf
1}_0\oplus{\bf 3}_0\oplus{\bf 3}_1\oplus{\bf 3}_{-1}$.

If a scalar state in the representation ${\bf 3}_1$ has energy 1, it
is a BPS state such that it must appear in the superconformal index.
Indeed, in the previous section we found four scalar BPS states with
energy 1 for the $OSp(M|2N)$ theory: $\bar A^{[a}\bar
A^{b]}\left|0,\cdots;0,\cdots\right>$ and $\bar A^{(a}_{1\hat 1}
\bar A^{b)}_{1\hat 1}\left|1,0,\cdots;1,0,\cdots\right>$. According
to \cite{Bashkirov:2010kz} as one varies a parameter, cohomology
classes appear and disappear in pairs so that members of the pair
have R-charge differing by 1 and energy, angular momentum differing
by 1/2. Since there is no spinor BPS state that has energy 1/2 and
$R$-charge 0, or energy 3/2 and $R$-charge 2, the above four scalar
states    are protected from the deformation; i.e., they still exist
in the undeformed theory.\footnote {This is why we examine the
lowest component of the stress tensor multiplet at first instead of
the $R$-currents themselves. By looking at the scalar component, it's
easier to argue that wanted states exist in the strong coupling
region.} The first state with vanishing GNO charges is in the
representation ${\bf 1}_1$ of $SU(2)\times U(1)_R$ and would be in
the representation $\bf 5$ of $Spin(5)_R$. What we are really
interested in are the next three. The next three states are in the
representation ${\bf 3}_1$ of $SU(2)\times U(1)_R$. Furthermore,
they must be in a representation of $Spin(5)_R$, the $R$-symmetry of
the undeformed theory. The multiplet ${\bf 3}_1$ must lie in a
representation of $Spin(5)_R$ having the highest weight (1,1)
because the multiplet ${\bf 3}_1$ contains BPS states with a weight
(1,1). Representation of $Spin(5)_R$ containing (1,1) as the highest
weight is the rank 3 antisymmetric tensor representation $\bf
10$. Therefore, we conclude that three scalar BPS states $\bar
A^{(a}_{1\hat 1} \bar A^{b)}_{1\hat
1}\left|1,0,\cdots;1,0,\cdots\right>$ are in the representation $\bf
10$ of $Spin(5)_R$ and that the undeformed $OSp(M|2N)$ theory has
the 3rd rank antisymmetric tensor multiplet $\bf 10$. Combined
with $\bf 5$ of the lowest scalar component of the $Spin(5)_R$,
these provide the lowest scalar component of the $Spin(6)_R$.

Once we obtain the needed scalar component, we can obtain
$R$-currents by acting superconformal generators. By acting
$Q^{(\alpha}_{[a}Q^{\beta)}_{b]}$ on the scalar component $\bf 10$,
we obtain vector states of spin 1 which are in the  representation
$\bf 5$ of $Spin(5)_R$.\footnote{We also have a $Spin(5)_R$ singlet
$\bf 1$ vector state of spin 1  due to the equivalence of the
2nd rank and 3rd rank antisymmetric representations of $Spin(5)_R$.
This is consistent with the conclusion for ${\cal N}=6$ theories of
\cite{Bashkirov:2011fr} that a ${\cal N}=6$ theory always has the
$U(1)$ global symmetry, which must be also true for our theories if
they indeed have ${\cal N}=6$ symmetry. One can check that  norms of
vector states ${\bf 1}$ and ${\bf 5}$  do not vanish, for example,
by decomposing the ${\cal N}=6$ stress tensor multiplet with respect
to the ${\cal N}=5$ subalgebra because the norms are completely
determined by the superconformal algebra alone. The ${\cal N}=6$
stress tensor multiplet contains the scalar component $\bf 15$ of $Spin(6)_R$,
which decomposes as ${\bf 15}={\bf 5}\oplus{\bf 10}$ under
$Spin(5)\subset Spin(6)_R$, and the vector components ${\bf
15}\oplus{\bf 1}$ of $Spin(6)_R$, which again decomposes as ${\bf
15}+{\bf 1}={\bf 10}\oplus{\bf 5}\oplus{\bf 1}$ under the $Spin(5)$. Thus,
any ${\cal N}=5$ theory having a scalar component $\bf 10$ of
$Spin(5)_R$ indeed has vector components ${\bf 5}\oplus{\bf 1}$ of
$Spin(5)_R$ with nonvanishing norm.} Since the vector states in $\bf
5$ have energy 2, operators corresponding to those vector states are
conserved currents by unitarity. We now have additional conserved
currents in the representation $\bf 5$ of $Spin(5)_R$ along with the
$R$-currents in the adjoint representation ${\bf 10}$ of
$Spin(5)_R$. They must fit into the adjoint representation of some
Lie group. In this case, it should be $Spin(6)$. The adjoint
representation of $Spin(6)$ decomposes under its subgroup $Spin(5)$
as ${\bf 15}={\bf 10}\oplus{\bf 5}$. Thus, ${\cal N}=5$
supersymmetry of the $OSp(M|2N)$ theory is enhanced to ${\cal N}=6$.
Note that this enhancement occurs only for $O(M)_{2k} \times
USp(2N)_{-k}$ with $k=1$. The BPS states transforming as ${\bf 3_1}$
having energy 1 at (\ref{states3}) exist only for $k=1$. For higher
$k>1$ one cannot have such states with energy 1 due to the Gauss
constraints which require higher energy states.

By slightly modifying the above argument, one can show that ${\cal
N}=5$ superconformal theory with $U(1)$ global symmetry leads to
${\cal N}=6$ superconformal theory. Note that $U(1)$ global current
belongs to a supermultiplet, whose lowest scalar component has
conformal dimension or energy
$\epsilon=1$ and 2nd rank antisymmetric tensor of $Spin(5)_R$, ${\bf
10}$. We already have the scalar component of rank 4 tensor of
$Spin(5)_R$ in the stress tensor multiplet, which transforms as $\bf 5$.
By applying $Q^{(\alpha}_{[a}Q^{\beta)}_{b]}$ on the multiplet $\bf
10$ again, we obtain vector states $\bf 5$ of $\epsilon=2$ while
acting on $\bf 5$ we have vector states $\bf 10$. Together they
transform as the adjoint representation of $SO(6)_R$ and we have ${\cal
N}=6$ superconformal theory. Note that the conserved current for the
$U(1)$ global symmetry exists apart from such $\bf 15 \,\,\,$
$R$-currents, which means that ${\cal N}=6$ theory still has that
$U(1)$ symmetry as a global symmetry.

\section{Large $N$ index for $\mathcal{N}=5$ theories and gravity index on  $AdS_4 \times
S^7/D_k$}

In this section, we will argue that the superconformal index for
$\mathcal{N}=5$ theory at large $N$ exactly matches the gravity index
on $AdS_4\times S^7/D_k$. Especially for $k=1$, $D_1=Z_4$ and the
exact match gives an additional evidence for the equivalence between
$OSp$ type $\mathcal{N}=5$ theory with $k=1$ and $U$ type
$\mathcal{N}=6$ theory with $k=4$ at large $N$.  At large $N$, the
difference between $OSp(2N|2N)$ and $OSp(2N+1|2N)$  is negligible
and two theories give the same index. The subtlety between $SO(N)$
and $O(N)$ is also negligible at  large $N$.

In the appendix we derive the superconformal index for
$SO(2N)_{2k}\times USp(2N)_{-k}$, which is given by
\begin{eqnarray}
&&I^{SO(2N)_{2k}\times USp(2N)_{-k}}
\\
&&= \sum_{\{n_i \}, \{\tilde{n}_i\}} \frac{ x^{\epsilon_0
}}{\textrm{(sym)}} \int \prod \frac{d\lambda_i
d\tilde{\lambda}_i}{(2\pi)^2} e^{2i k \sum_i (n_i \lambda_i -
\tilde{n}_i \tilde{\lambda}_i)} \exp \big{[} \sum_{n=1}^\infty
\frac{1}n  I_{sp}(x^n, e^{i n \lambda_i},e^{in
\tilde{\lambda}_i})\big{]}.\nonumber
\end{eqnarray}
with the single letter index $I_{sp}$ is
\begin{eqnarray}
&&I_{sp} (x, e^{i \lambda_i}, e^{i \tilde{\lambda}_i}) =
f(x)\sum_{\pm}\sum_{i,j}  \big{(} e^{\pm i(\lambda_i -
\tilde{\lambda}_j)} x^{|n_i- \tilde{n}_j|}+ e^{\pm i(\lambda_i +
\tilde{\lambda}_j)} x^{|n_i+ \tilde{n}_j|}  \big{)}  \nonumber
\\
&&- \sum_{\pm} \sum_{i<j} \big{(} e^{ \pm i (\tilde{\lambda}_i -
\tilde{\lambda}_j)} x^{|\tilde{n}_i - \tilde{n}_j|}+ e^{ \pm i (
\tilde{\lambda}_i +\tilde{ \lambda}_j)} x^{|\tilde{n}_i +\tilde{
n}_j|} \big{)} - \sum_{\pm} \sum_{i } e^{\pm 2 i \tilde{\lambda}_i}
x^{|2\tilde{n}_i|} \nonumber
\\
&&- \sum_{\pm} \sum_{i<j} \big{(} e^{ \pm i ( \lambda_i -
\lambda_j)} x^{|n_i - n_j|}+ e^{ \pm i ( \lambda_i + \lambda_j)}
x^{|n_i + n_j|} \big{)}\;, \;\textrm{where } f(x):=
\frac{2x^{\frac{1}2} }{1+x} \;. \label{single letter indexmain}
\end{eqnarray}
 Now we will take
the large $N$ limit on the superconformal index. From eq. (\ref{single
letter indexmain}), we will use a similar large $N$ analysis technique
used in \cite{Kim:2009wb} (see also \cite{Gang:2011xp}). To take
the large $N$ limit, we  first introduce ($n=0,1,\ldots$)
\begin{align}
\rho_n = \sum_{j=N_1+1}^{N} e^{ in \lambda_j}+e^{ -in \lambda_j},
\quad \chi_n = \sum_{j=N_2+1}^{N} e^{ in \tilde{\lambda}_j}+e^{- in
\tilde{\lambda}_j}\;.
\end{align}
We assume that first $N_1$ ($N_2$) monopole fluxes for $SO(2N)$
($USp(2N)$)  are non-zero and the rest   are all zero. In terms of
$(\rho_n, \chi_n)$ variables,
\begin{eqnarray}
&&\exp \big{[} \sum_{n=1}^\infty \frac{1}n  I_{sp}(x^n, e^{i n
\lambda_i},e^{in \tilde{\lambda}_i})\big{]} \nonumber
\\
&&=\exp[- \sum_{p\textrm{ odd}}^\infty \frac{1}{2p} \rho_p^2 -
\sum_{p\textrm{ even}}^\infty \frac{1}{2p} (\rho_p^2-2 \rho_p) -
\sum_{p\textrm{ odd}}^\infty \frac{1}{2p} \chi_p^2 - \sum_{p\textrm{
even}}^\infty \frac{1}{2p} (\chi_p^2 + 2 \chi_p) +\sum_p
\frac{1}{p}f(x^p) \rho_p \chi_p ] \nonumber
\\
&&\quad \times \exp [\sum_{p=1}^\infty \frac{1}p \rho_p \big{(}
\sum_{i=1}^{N_2} x^{p |\tilde{n}_i|}(e^{i p \tilde{\lambda}_i}+e^{-i
p \tilde{\lambda}_i})f(x^p)- \sum_{i=1}^{N_1} x^{p|n_i|}(e^{ip
\lambda_i}+e^{-ip \lambda_i}) \big {)} ] \nonumber
\\
&&\quad \times \exp [\sum_{p=1}^\infty \frac{1}p \chi_p \big{(}
\sum_{i=1}^{N_1} x^{p |n_i|}(e^{i p \lambda_i}+e^{-i p
\lambda}_i)f(x^p)- \sum_{i=1}^{N_2} x^{p|\tilde{n}_i|}(e^{ip
\tilde{\lambda}_i}+e^{-ip \tilde{\lambda}_i}) \big {)} ]  \nonumber
\\
&& \quad \times  \exp \big{[} \sum_{n=1}^\infty \frac{1}n
 I^{OSp(2N_1|2N_2)}_{sp}(x^n, e^{i n \lambda_i},e^{in
\tilde{\lambda}_i})\big{]} \;.
\end{eqnarray}
where $I_{sp}^{OSp(2N_1|2N_2)}$ denotes the single letter index for
$SO(2N_1)\times USp(2N_2)$ theory, which is the same as
\eqref{single letter indexmain} except that the index $i$ in
$(\lambda_i, n_i)$ (or $(\tilde{\lambda}_i,\tilde{ n}_i)$)  runs
from 1 to $N_1$ (or $N_2$). In the large $N$ limit, the holonomy
variable integrations can be replace by integration of $(\rho_n,
\chi_n)$ variables
\begin{eqnarray}
\int \prod_{i=1}^N \frac{d\lambda_i d\tilde{\lambda}_i}{(2\pi)^2} \;
\rightarrow \; \int \prod d \rho_n d \chi_n \;.
\end{eqnarray}
The infinite dimensional integral for $(\rho_n, \chi_n)$ is gaussian
and can be easily performed. Doing the gaussian integration and
simplifying the formula, we finally get
\begin{eqnarray}
I^{O(2N)_{2k}\times Sp(2N)_{-k}}_{N\rightarrow \infty} (x)=I^{(0)}
(x) I^\prime (x) \;.
\end{eqnarray}
$I^{(0)}(x)$ comes from zero monopole fluxes.
\begin{eqnarray}
&&I^{(0)}(x) = \prod_{n=1}^\infty \frac{ 1}{\sqrt{1-f^2 (x^n)}}\exp
\big{[}- \frac{f(x^{2n})}{2n (1+f (x^{2n}))} \big{]}\;.
\label{I^{(0)}}
\end{eqnarray}
$I^{\prime} (x)$ is given by
\begin{eqnarray}
I^\prime(x)= \sum_{\{n_i \}, \{\tilde{n}_i\}} \frac{ x^{\epsilon_0
}}{\textrm{(sym)}} \int \prod (\frac{d\lambda_i}{2\pi})
(\frac{d\tilde{\lambda}_i}{2\pi}) e^{2i k \sum_i (n_i \lambda_i -
\tilde{n}_i \tilde{\lambda}_i)} \exp \big{[} \sum_{n=1}^\infty
\frac{1}n  I^\prime_{sp}(x^n, e^{i n \lambda_i},e^{in
\tilde{\lambda}_i})\big{]}\;,\nonumber
\end{eqnarray}
where
\begin{eqnarray}
&&I^\prime_{sp} (x, e^{i \lambda_i}, e^{i \tilde{\lambda}_i}) =
f(x)\sum_{\pm}\sum_{i,j}   e^{\pm i(\lambda_i - \tilde{\lambda}_j)}
(x^{|n_i- \tilde{n}_j|}-x^{|n_i|+|n_j|})  \nonumber
\\
&&- \sum_{\pm} \sum_{i<j} e^{ \pm i ( \lambda_i - \lambda_j)}(
x^{|n_i - n_j|}-x^{|n_i| +| n_j|} )- \sum_{\pm} \sum_{i<j}e^{ \pm i
(\tilde{\lambda}_i - \tilde{\lambda}_j)} (x^{|\tilde{n}_i -
\tilde{n}_j|}-x^{|\tilde{n}_i| +| \tilde{n}_j|})\;. \nonumber
\end{eqnarray}
By comparing the above formulae with the large $N$ index formulae for
$U(N)_k \times U(N)_{-k}$ theory   in \cite{Kim:2009wb}, one can see
that
\begin{equation}
I^{\prime}(x) = I_{N\rightarrow \infty  :(+)}^{U(N)_{2k}\times
U(N)_{-2k}} (x)
\end{equation}
except the Casimir energy $\epsilon_0$. The difference between
Casimir energies  in two formulae  is $\sum n_i - \sum \tilde{n}_i$.
However,  as already noticed in \cite{Kim:2009wb},
 only monopole fluxes satisfying  $\sum n_i  = \sum \tilde{n}_i$ contribute to the large $N$ index and
 thus the difference in $\epsilon_0$ vanishes at large $N$.  In \cite{Kim:2009wb} the large $N$ index for
 $U(N)_{k}\times U(N)_{-k}$ theory is shown to be factorized into three factors, contribution from zero
 monopole flux and contributions from positive/negative monopole fluxes only.
  $ I_{N\rightarrow \infty  :(+)}^{U(N)_{k}\times U(N)_{-k}}$  denote the  factor from positive
  fluxes only which is actually the same as the factor from negative fluxes only.
  Note that for ${\cal N}=5$ theories we only need to consider contributions from positive monopole fluxes only, thanks
  to Weyl symmetries. Thus we found following
  relation between large index for $\mathcal{N}=5$ and $\mathcal{N}=6$ theories.
\begin{equation}
I^{O(2N)_{2k}\times USp(2N)_{-k}}_{N\rightarrow \infty} (x)=I^{(0)}
(x) I_{N\rightarrow \infty :(+)}^{U(N)_{2k}\times U(N)_{-2k}} (x)\;.
\label{large N index}
\end{equation}

To show the equality between the large $N$ index and  gravity index
on $S^7/D_k$,  we will assume that the large $N$ index for $U(N)_k
\times U(N)_{-k}$ is the same as gravity index on $S^7/Z_k$. This is
not yet proved but checked in various sectors \cite{Kim:2009wb} and
believed to be true. Two generators $\alpha,\beta$ of the dihedral
group $D_k$ act on $S^7$ as (see section 3 in \cite{Choi:2008za})
\begin{align}
\alpha := \exp (\frac{2 \pi i }{k} J_3), \quad \beta :=\exp(\pi i
J_2).
\end{align}
$J_{1,2,3}$ are three generators (with normalization $[J_i , J_j]=
i\epsilon_{ijk} J_k$) of $SU(2) \simeq SO(3)$ in $SO(5)\times SO(3)
\subset SO(8)$, isometry group on $S^7$. $J_3$ can be identified
with the baryonic $U(1)_b$ symmetry in $U(N|N)$ theories.  Graviton
spectrum on $S^7/D_k$ can obtained by keeping only  $D_k$ invariant
states in graviton spectrum on $S^7$. In terms of the $SU(2)$
charges, the $D_k$ invariant states can be divided by two types
\begin{align}
&\textrm{type $I$ : }  |\ell  , m=0\rangle, \quad \ell \in
2\mathbb{Z}\;, \nonumber
\\
&\textrm{type $II$ : }  (|\ell,m  \rangle + |\ell,-m \rangle ),
\quad \ell\in \mathbb{Z}, \quad m \in  k \mathbb{Z}_+.
\end{align}
$\mathbb{Z}$ and $\mathbb{Z}_+$ denote the set of integers and of
positive integers respectively. States $|\ell,m\rangle$ are
represented by its total angular momentum $\ell$ and angular
momentum in the 3rd direction, $m=J_3$.   In the second line, we
used the fact that $\{ J_3, \beta \} =0$ and $\beta^2 =1$ in
integer-spin representations of $SU(2)$. Gravity index from
gravitons of type $I$ are analyzed in \cite{Choi:2008za} and it
gives  exactly the same factor $I^{(0)}(x)$ in \eqref{I^{(0)}}.
Gravitons of  type $II$  can be thought as $Z_2$ invariant gravitons
on $S^7/Z_{2k}$ with non-zero $U(1)_b$ charge where the $Z_2$ flips
the sign of $U(1)_b$ charge. Thus, the gravity index from gravitons
of  type $II$ is $I_{S^7/Z_{2k}:(+)}$, gravity index from single
graviton with positive   $U(1)_b$ charge on $ S^7/Z_{2k}$. By
assuming the equality between large $N$ index for $U(N|N)$ theory and
gravity index on $ S^7/Z_k$, $I_{S^7/Z_{2k}:(+)}$ is nothing but
$I_{N\rightarrow \infty :(+)}^{U(N)_{2k}\times U(N)_{-2k}}$. In
summary, we found that
\begin{eqnarray}
&I_{S^7/D_k} = &I^{(0)} (\textrm{from gravitions of type $I$})
\nonumber
\\
&&\times I_{N\rightarrow \infty  :(+)}^{U(N)_{2k}\times
U(N)_{-2k}}(\textrm{from gravitions of type $II$})\;.
\end{eqnarray}
This perfectly matches the large $N$ index in eq.~\eqref{large N
index} for $O(2N)_{2k}\times USp(2N)_{-k}$ theory.
Note that for $k=1$, $D_1=Z_4$ so that  the large $N$ index
 for $O(2N)_{2}\times USp(2N)_{-1}$ theory is the same as the gravity index on $AdS_{S^7/Z_{4}}$,
 which in turn is the same as the large $N$ index
 for $U(N)_{4}\times U(N)_{-4}$ theory.
\section*{Acknowledgments}

S. Nagaoka acknowledges the Korea Ministry of Education, Science and
Technology (MEST) for the support of the Young Scientist Training
Program at the Asia Pacific Center for Theoretical Physics (APCTP).
This work was supported by the National Research Foundation of Korea
(NRF) grant funded by the Korea government (MEST)  with the Grants
No.~2009-0085995, 2012-046278 (JP) and 2005-0049409 (JP) through the
Center for Quantum Spacetime (CQUeST) of Sogang University. JP is
also supported by the POSTECH Basic Science Research Institute Grant
and appreciates APCTP for its stimulating environment for research.

\appendix
\section{Superconformal index formulae for the 3-d $O(M) \times USp(2N)$ gauge theory}

The index for $SO(2N)_{2k}\times USp(2M)_{-k}$ theory can be written
in the following form.
\begin{eqnarray}
&&I^{SO(2N)_{2k}\times USp(2N)_{-k}}
\\
&&= \sum_{\{n_i \}, \{\tilde{n}_i\}} \frac{ x^{\epsilon_0
}}{\textrm{(sym)}} \int \prod \frac{d\lambda_i
d\tilde{\lambda}_i}{(2\pi)^2} e^{2i k \sum_i (n_i \lambda_i -
\tilde{n}_i \tilde{\lambda}_i)} \exp \big{[} \sum_{n=1}^\infty
\frac{1}n  I_{sp}(x^n, e^{i n \lambda_i},e^{in
\tilde{\lambda}_i})\big{]}.\nonumber
\end{eqnarray}
Here $(\lambda_i , n_i)|_{i=1,\ldots,N}$ and $(\tilde{\lambda}_i,
\tilde{n}_i)|_{i=1,\ldots,M}$ are (holonomy variables, monopole
fluxes)  for $SO(2N)$ and $USp(2N)$ respectively. Using an Weyl
action of gauge group,  one can take all the monopole fluxes be
non-negative, $n_i, \tilde{n}_i \geq 0$. $\textrm{(sym)}$ denote a
symmetry factor, order (number of elements) of Weyl group for the
unbroken gauge group by monopole fluxes. The Casimir energy
$\epsilon_0$ is given by
\begin{align}
\epsilon_0 = \sum_{i,j}|n_i -\tilde{n}_j| - \sum_{i<j}^N|n_i -
n_j|-\sum_{i<j}^M|\tilde{n}_i - \tilde{n}_j|+\sum_{i}n_i -\sum_{i}
\tilde{n}_i\;.
\end{align}
The single letter index $I_{sp}$ is
\begin{eqnarray}
&&I_{sp} (x, e^{i \lambda_i}, e^{i \tilde{\lambda}_i}) =
f(x)\sum_{\pm}\sum_{i,j}  \big{(} e^{\pm i(\lambda_i -
\tilde{\lambda}_j)} x^{|n_i- \tilde{n}_j|}+ e^{\pm i(\lambda_i +
\tilde{\lambda}_j)} x^{|n_i+ \tilde{n}_j|}  \big{)}  \nonumber
\\
&&- \sum_{\pm} \sum_{i<j}^M \big{(} e^{ \pm i (\tilde{\lambda}_i -
\tilde{\lambda}_j)} x^{|\tilde{n}_i - \tilde{n}_j|}+ e^{ \pm i (
\tilde{\lambda}_i +\tilde{ \lambda}_j)} x^{|\tilde{n}_i +\tilde{
n}_j|} \big{)} - \sum_{\pm} \sum_{i }^M e^{\pm 2 i
\tilde{\lambda}_i} x^{|2\tilde{n}_i|} \nonumber
\\
&&- \sum_{\pm} \sum_{i<j}^N \big{(} e^{ \pm i ( \lambda_i -
\lambda_j)} x^{|n_i - n_j|}+ e^{ \pm i ( \lambda_i + \lambda_j)}
x^{|n_i + n_j|} \big{)}\;, \;\textrm{where } f(x):=
\frac{2x^{\frac{1}2} }{1+x} \;. \label{single letter index}
\end{eqnarray}
We should consider the additional projection for $Z_2$ element of
$O(2N)$ not belonging to $SO(2N)$ group. We choose the specific
$Z_2$ action,
\begin{equation}
Z_2= \left(\begin{array}{cccc}
1&&&\\
&-1&&\\
&&1&\\
&&&\ddots
\end{array}\right).
\end{equation}
Under this $Z_2$ action, the eigenvalues of the holonomy and the
monopole are projected into
\begin{equation}
e^{\pm i\lambda_1}\rightarrow\pm1,~~~~~~\pm n_1\rightarrow0.
\end{equation}
The other variables are not affected.

 The single letter index for
$SO(2N+1)_{2k}\times USp(2M)_{-k}$ is given by
\begin{eqnarray}
&&I_{sp} (x, e^{i \lambda_i}, e^{i \tilde{\lambda}_i}) =
f(x)\sum_{\pm}\sum_{i,j}  \big{(} e^{\pm i(\lambda_i -
\tilde{\lambda}_j)} x^{|n_i- \tilde{n}_j|}+ e^{\pm i(\lambda_i +
\tilde{\lambda}_j)} x^{|n_i+ \tilde{n}_j|} +  e^{\pm i
\tilde{\lambda}_j} x^{|\tilde{n}_j|}\big{)}  \nonumber
\\
&&- \sum_{\pm} \sum_{i<j}^M \big{(} e^{ \pm i (\tilde{\lambda}_i -
\tilde{\lambda}_j)} x^{|\tilde{n}_i - \tilde{n}_j|}+ e^{ \pm i (
\tilde{\lambda}_i +\tilde{ \lambda}_j)} x^{|\tilde{n}_i +\tilde{
n}_j|} \big{)} - \sum_{\pm}^M \sum_{i } e^{\pm 2 i
\tilde{\lambda}_i} x^{|2\tilde{n}_i|} \nonumber
\\
&&- \sum_{\pm} \sum_{i<j}^N \big{(} e^{ \pm i ( \lambda_i -
\lambda_j)} x^{|n_i - n_j|}+ e^{ \pm i ( \lambda_i + \lambda_j)}
x^{|n_i + n_j|} \big{)}-\sum_{\pm} \sum_{i}^N e^{ \pm i
\lambda_i}x^{|n_i|}\;, \;\textrm{where } f(x):= \frac{2x^{\frac{1}2}
}{1+x} \;. \nonumber
\end{eqnarray}
Let us turn to $O(2N+1)$ theory.
%With facts that the weights of the
%fundamental representation are $\pm \epsilon_i$ where $i=1,\cdots,N$
%and that the roots of $O(2N+1)$ are $\pm \epsilon_i$ and
%$\pm\epsilon_i\pm\epsilon_j$ where $i,j=1,\cdots,N$ and $i\neq j$,
In this case, we choose $Z_2$ action,
\begin{equation}
Z_2= \left(\begin{array}{cccc}
1&&&\\
&\ddots&&\\
&&1&\\
&&&-1
\end{array}\right),
\end{equation}
an eigenvalue 1 of the holonomy in the fundamental representation is
projected by
\begin{equation}
1\rightarrow-1
\end{equation}
while the others are not influenced. Furthermore, eigenvalues
$e^{\pm i\lambda_i}$ of the holonomy in the adjoint representation
are projected by
\begin{equation}
e^{\pm i\lambda_i}=e^{\pm i\lambda_i}\cdot1\rightarrow e^{\pm
i\lambda_i}\cdot(-1)
\end{equation}
while the others, which are in the form of $e^{i(\pm \lambda_i\pm
\lambda_j)}=e^{\pm i\lambda_i}\cdot e^{\pm i\lambda_j}$, are not
affected, either.

\end{document}